# Thiol-amine co-solvents aided direct synthesis of ZnTe thin films by spin coating for low cost optoelectronic applications

Sheikh Noman Shiddique[1], Syeda Samiha Nushin[1], Bipanko Kumar Mondal[2], Ahnaf Tahmid Abir[1], Md. Mahbubor Rahman[3], Mainul Hossain[4], and Jaker Hossain[1*]

[1]*Solar Energy Laboratory, Department of Electrical and Electronic Engineering, University of Rajshahi, Rajshahi 6205, Bangladesh.*

[2]*Department of Electrical and Electronic Engineering, Begum Rokeya University, Rangpur, Rangpur 5400, Bangladesh.*

[3]*Department of Chemistry, University of Rajshahi, Rajshahi 6205, Bangladesh.*

[4]*Department of Electrical and Electronic Engineering, University of Dhaka, Dhaka-1000, Bangladesh.*

**Abstract**

Zinc telluride (ZnTe) thin films have special semiconducting characteristics that make them very promising for a broad range of optoelectronic applications. In this work, a novel approach for synthesizing ZnTe thin films by spin coating technique is followed using a unique solution process with ZnTe directly dissolving in thiol-amine co-solvents. Thin films are synthesized on glass substrates and air annealed at 250-350 °C. The polycrystalline phase of ZnTe is revealed through the X-ray diffraction (XRD) study. The scanning electron microscopy (SEM) is used to observe the evolution of surface smoothness with annealing temperature. Moreover, elemental compositions of ZnTe thin film have been determined by energy dispersive spectroscopy (EDS) study. FTIR spectroscopy reveals that ZnTe has been successfully synthesized as confirmed by the characteristic peaks in the spectrum of 750-1000 cm$^{-1}$. Optical properties of the ZnTe thin films have been investigated using UV-vis spectroscopy. The transmittance of the films increases with annealing temperature. Furthermore, the optical bandgaps of the films of 2.92, 2.84, and 2.5 eV have been found at 250, 300, and 350 °C annealing temperatures, respectively. These results suggest that controlling the annealing environment serves as a valuable strategy for tailoring the ZnTe film properties to meet specific application requirements. These results reveal that spin coated ZnTe thin films are attractive ones for various applications in optoelectronic devices such as solar cells and photodetectors.

## 1. Introduction

Zinc telluride (ZnTe) has been the focus of noteworthy studies lately because of its exceptional semiconducting characteristics. These attributes have allowed ZnTe to be used in many different applications such as thin-film solar cells, light-emitting diodes, photodetectors, and thin-film transistors [1-4]. ZnTe is a vital member of the II-VI metal chalcogenide group containing a direct bandgap of 2.26 eV [5]. Moreover, this bandgap might be adjusted from 1.8 to 3.1 eV to raise the practical applicability of ZnTe [6]. The wider bandgap, low electron affinity of 3.53 eV, and p-type conductivity render it the perfect window or buffer layer in CdSe and CdTe-based photovoltaic cells [7]. In addition, it has also been used as a back contact layer in CdTe-based solar cells due to its lower resistivity [8]. The majority of electrical research points it to p-type conductivity; nevertheless, n-type ZnTe can be created by doping it with Al, Cl, and Sn [9-10]. The n-type ZnTe would be a suitable substitute for toxic, and hazardous CdS as a window layer in solar cells' applications. ZnTe has demonstrated its promise in recent times by employing as back surface field (BSF) with CuInS$_2$ absorber layer [6].

However, the synthesized crystalline phase of ZnTe could be found in a cubic or hexagonal structure. The zinc-blend-like cubic structure with 6.103 Å lattice constant of ZnTe belongs to the $F\bar{4}3m$ space group, while the hexagonal structure with a lattice of a = 4.045 and c = 9.342 Å corresponds to the P3$_1$ space group [11-12].

Thus far, numerous fabrication techniques have already been utilized to grow ZnTe films including hydrothermal [13], molecular beam epitaxy (MBE) [9], closed space sublimation (CSS) [14], magnetron sputtering [15], metallo-organic chemical vapor deposition (MOCVD) [16], thermal evaporation [17], electrodeposition [10], metallo-organic vapor phase epitaxy (MOVPE) [18], hot-wall evaporation [19] and chemical bath deposition [20] etc. To the best of our knowledge, ZnTe thin films have not yet been synthesized via the spin coating solution-processed technique. Especially compared to certain other thin-film deposition methods, the spin coating method is very easy to use and reasonably priced. It doesn't call for sophisticated equipments. Furthermore, it is possible to precisely control the thickness of the deposited film by varying both the coating solution's viscosity and rotating speed in this method. Hence, the facile spin coating method has been used in this work to deposit ZnTe thin films.

In recent years, bulk V$_2$VI$_3$ (V = As, Bi, Sb; VI = Te, Se, S) group and several chalcogenide compounds have been shown to dissolve in a basic thiol-amine co-solvents system [21-22]. The CdS, CdTe, γ-In$_2$Se$_3$, and β-In$_3$Se$_2$ thin films have already been fabricated using this route [23-26]. Still, it is rare to come across any literature detailing the production of ZnTe thin films using a thiol-amine co-solvents solution that dissolves ZnTe powder directly.

Therefore, in this article, we have synthesized ZnTe thin films utilizing ethylene-di-amine and 1,2 ethane-di-thiol co-solvents solution by spin coating deposition method on to glass substrates. The fabricated ZnTe films have been annealed at various conditions and characterized by X-ray diffraction (XRD), scanning electron microscopy (SEM), Energy-dispersive X-ray spectroscopy (EDS), Fourier-transform Infra-red (FTIR) Spectroscopies to analysis crystalline phase, surface morphology, elemental composition and local bonds to unveil its potential.

## 2. Experimental details

### *2.1 Materials and Precursor Solution Preparation*

ZnTe solution has been prepared by using high-purity ZnTe granular powder (99.99% purity), ethylene-diamine, and 1,2-ethane-dithiol co-solvents purchased from Sigma Aldrich. The methodical phases in the preparation process, as shown in Figure 1(a), ensured a precise formulation of the solution. Initially, 1w% of ZnTe is dissolved in a mixture of ethylene-diamine and 1,2-ethanedithiol co-solvents in a volume ratio of 9:1. A magnetic stirrer has been used to agitate the mixture in a closed container at 350 rpm and 50 °C. After agitating the mixture for about 26 hours, a dark brown solution is resulted. The solution is then stored in an air-tight container to prevent any additional reactions with atmospheric air.

### *2.2 Thin film deposition*

The schematic representing the sequential steps involved in cleaning the substrate and depositing zinc telluride thin films is shown in Figure 1(b). This schematic serves as a visual guide, outlining the steps required for achieving precise and efficient deposition of the ZnTe thin films on the substrate surface. Initially, detergent is used to clean the (1×1)-inch glass substrates. Subsequently, a piranha solution composed of $H_2SO_4$ and $H_2O_2$ in a volume ratio of 5:1 is utilized for additional cleaning to eliminate any remaining glass surface impurities. During this step, the substrate needs to be immersed in the piranha solution for 30 minutes. Thereafter, distilled water comes at its role to rinse the substrate. Following the cleaning procedures, the substrate is dried at 100 °C for 10 minutes.

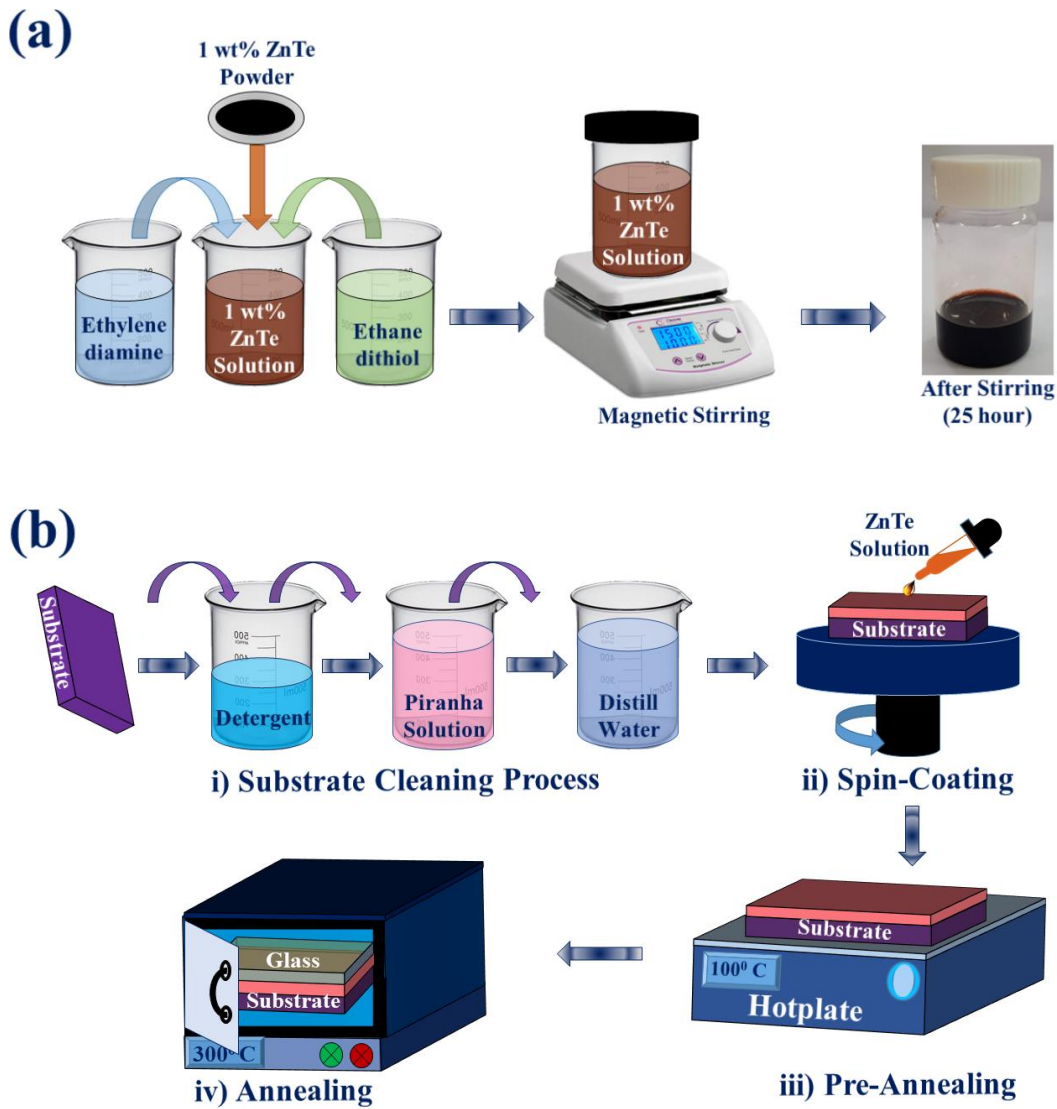

**Figure 1:** **(a)** The steps of ZnTe-thiol-amine solution preparation and **(b)** Diagram showing the procedures involved in cleaning the substrate and depositing the ZnTe thin films.

Next, the prepared substrate has been positioned onto a spin coater, a rotating platform. Then, the ZnTe solution has been carefully dispense onto the substrate's surface using a 0.45-micron filter syringe. The solution is then spin-coated onto the substrate to create the thin film. The rotational speed has been set to 1000 rpm for 30 seconds, followed by a gradual decrease to 500 rpm over 10 seconds. Afterward, the films have been pre-annealed for ten minutes at 100 °C to eliminate any remaining solvent. Subsequently, the films have been subjected to post-annealing at three different temperatures of 250, 300, and 350 °C, each for 10 minutes under stressed conditions to minimize oxidation [23,26]. The mechanical stress has been applied by clamping the film between two microscope slides, each (1-1.2) mm thick using stainless steel clips secured to the edges.

## 2.3 Thin film characterization

The ZnTe film thickness has been determined by analyzing the cross-sectional SEM images. An average thickness of synthesized ZnTe films has been found to be 190 nm. Structural analysis of these films are performed using the X-ray Diffractometer (XRD; GIXRD/XRR: ARL EQUINOX 1000), resolution: wavelength, $\lambda = 1.54$ Å and scans in the range of $2\theta = (5°-100°)$. The surface nature of the ZnTe films is assessed visually using scanning electron microscopy (SEM) (model: JEOL JEM- 6500F). EDS study has done by SEM (model: JSM-IT800(IS)). Furthermore, a PerkinElmer (Spectrum 100) associates to aggregate the FTIR spectra within the wavenumber limit of 225 to 4000 $cm^{-1}$ to clarify the chemical composition of the synthesized ZnTe films. Finally, optical features are characterized by examining the transmission spectra of the ZnTe films by using a spectrophotometer (UV-1900i PC, Shimadzu Corporation, Japan).

## 3. Results and Discussions

### *3.1 XRD study*

The XRD patterns in Figure 2 indicates that the ZnTe thin films display a polycrystalline structure across all annealing temperatures of 250-350 °C. For annealing temperature of 250 °C, the peaks detected at 2θ = 27.6°, 38.4°, 40.5°, 43.6°, 49.6° and 57.02° are agreed with the zinc blende (ZB) (cubic) structure with planes (111), (103), (220), (012), (311) and (021), respectively. Whereas at 300 °C, in addition to the peak at 49.6° and 57.02°, a new peak at 65.9° is appeared indicating (331) plane of ZB structure of ZnTe. The observed shift in the relative intensity of the ZnTe peak indicates a change in the orientation of grains, notably from (311), and (021) planes to (331), where they combine with the pre-existing grains are aligned along the (331) planes. Subsequent annealing is anticipated to result in an enlargement of grain size [27]. Further increase in annealing temperature to 350 °C, shows the consistence behavior like 250 °C. All these planes indicate ZB structure are corresponding to the (JCPDS Card No. 15-0746, 80-0022 and 83-0967) [28]. The appearance of distinct and exceptionally narrow diffraction peaks indicates that the manufactured ZnTe thin film has a high degree of crystalline quality. Significantly, the prominence of the (111) peak suggests a directional preference in ZnTe growth along the c-axis. Although ZnTe usually takes on the form of cubic zinc-blende (ZB) crystals [29], hexagonal crystals (WZ) are another possible form. Conventionally, the bulk ZB structure is more energetically favorable compared to WZ [30]. Peaks corresponding to ZB structure in the synthesized ZnTe thin film are shown in Figure 2.

Additionally, some weak peaks are observed at 2θ values of 23.01° and 72.09°, which are indicating the (100) and (122) orientations, respectively. These peaks are consistent with the hexagonal crystal structure of pure Te according to the (JCPDS Card No. 01-078-2312). Significantly, no distinctive impurity peaks are observed.

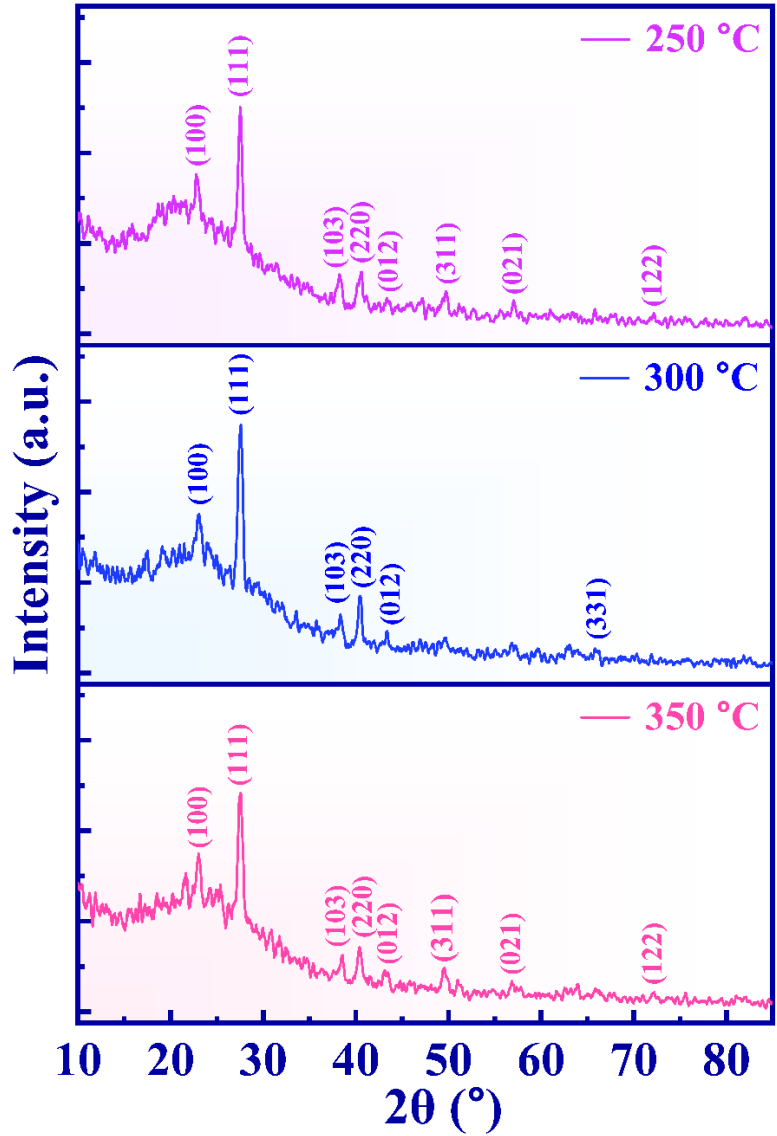

**Figure 2:** The XRD spectra of ZnTe thin films annealed at various temperatures in the span of 250-350 °C.

The structural features such as crystal size (D), density of dislocation (δ), and strain (ε) are obtained from the XRD study to gain insights into the crystallographic properties. The Debye-Scherrer's equation helps to compute the crystallite size (D) [31]

$$D = \frac{0.94\lambda}{\beta \cos\theta} \qquad (1)$$

Wherein, λ represents the wavelength of X-ray expressed in nm, β denotes the FWHM presented in radians, and θ signifies the Bragg angle of the respective XRD peak.

The following formula is essential to calculate the dislocations (δ), that denotes the dislocation lines per unit volume within the crystal [32]

$$\delta = \frac{1}{D^2} \qquad (2)$$

Strain (ε) is derived from the slope of the plot depicting βcosθ against sinθ using the following relation [33]

$$\beta = \frac{\lambda}{D\cos\theta} - \varepsilon \tan\theta \qquad (3)$$

Table 1 reveals a range of structural values for the synthesized ZnTe thin films. Where, crystallite sizes span from 22.7 to 115.8 nm, dislocation densities (δ) from $0.74 \times 10^{14}$ to $19.4 \times 10^{14}$ lines/m², and strain (ε) from $0.22 \times 10^{-3}$ to $57.2 \times 10^{-3}$ lines$^{-2}$m$^{-4}$. Notably, there is a discernible trend where an increase in crystallite size (D) correlates with a decrease in dislocation density (δ).

**Table 1:** A summary of the calculations for crystallite size, dislocation density and strain of ZnTe thin films.

| Annealing Temperature (°C) | Planes (hkl) | FWHM, β (rad × 10³) | Crystallite size, D(nm) | Dislocation density, δ(lin/m²) × 10¹⁴ | Strain, ε (lin⁻² .m⁻⁴) ×10⁻³ |
|---|---|---|---|---|---|
| 250 | (100) | 0.5904 | 28.700 | 12.140 | 3.512 |
|  | (111) | 0.2952 | 57.936 | 2.979 | 1.936 |
|  | (103) | 0.3936 | 44.683 | 5.008 | 0.665 |
|  | (220) | 0.1968 | 89.961 | 1.235 | 1.468 |
|  | (012) | 0.5904 | 30.299 | 10.892 | 56.896 |
|  | (311) | 0.5904 | 30.993 | 10.409 | 0.229 |
|  | (021) | 0.2952 | 64.028 | 2.439 | 20.651 |
|  | (122) | 0.1968 | 104.423 | 0.917 | 1.557 |
| 300 | (100) | 0.5904 | 28.708 | 12.133 | 3.233 |
|  | (111) | 0.1476 | 115.886 | 0.744 | 0.994 |
|  | (103) | 0.5904 | 29.783 | 11.273 | 0.926 |
|  | (220) | 0.2952 | 59.960 | 2.781 | 2.121 |
|  | (012) | 0.7872 | 22.702 | 19.402 | 41.874 |
|  | (331) | 0.5904 | 33.539 | 8.889 | 4.625 |
| 350 | (100) | 0.3936 | 43.061 | 5.392 | 2.179 |
|  | (111) | 0.2952 | 57.917 | 2.981 | 1.797 |
|  | (103) | 0.2952 | 59.593 | 2.815 | 0.551 |
|  | (220) | 0.3936 | 44.954 | 4.948 | 2.676 |
|  | (012) | 0.7872 | 22.699 | 19.407 | 39.842 |
|  | (311) | 0.5904 | 30.985 | 10.415 | 0.384 |
|  | (021) | 0.4920 | 38.382 | 6.787 | 57.297 |
|  | (122) | 0.5904 | 34.793 | 8.260 | 4.850 |

*3.2 SEM study*

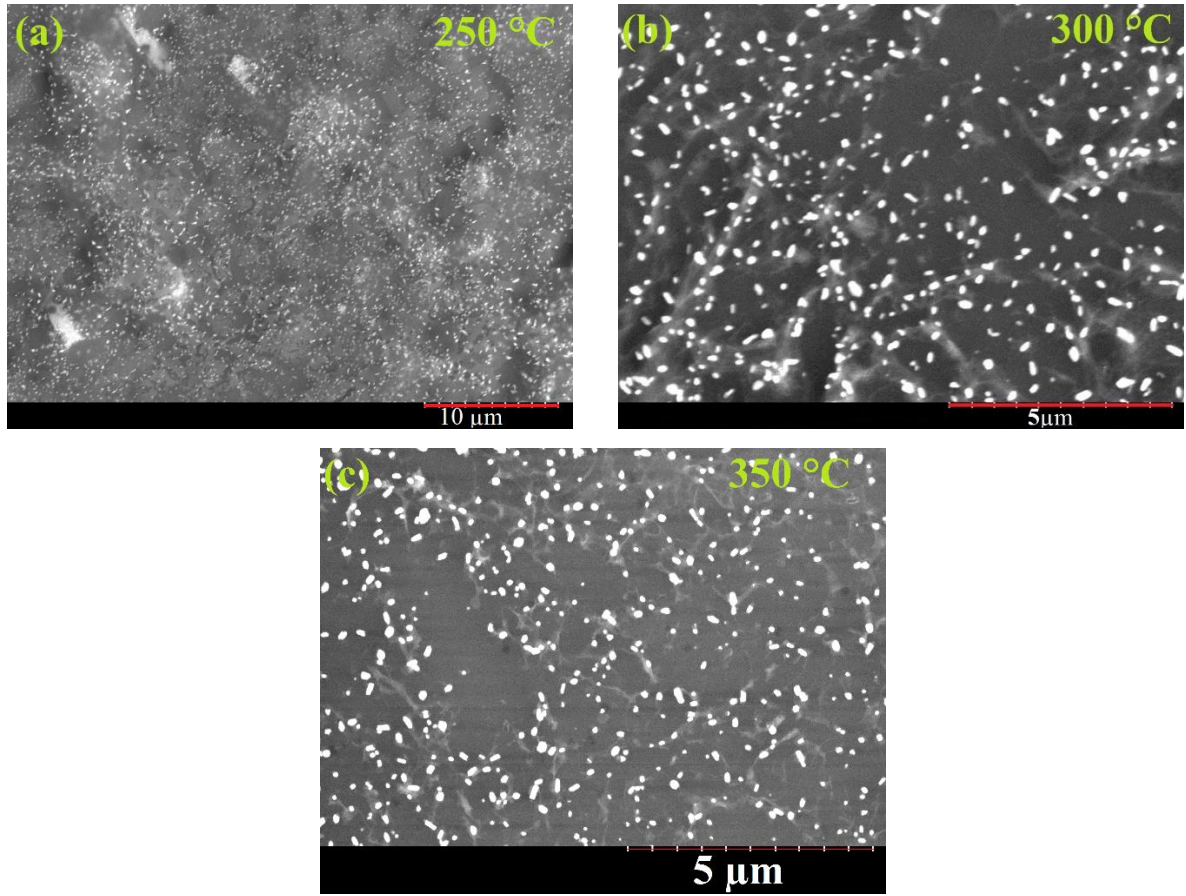

**Figure 3:** The SEM micrographs of spin-coated ZnTe thin films annealed at various temperatures.

ZnTe thin films undergo surface feature analysis using SEM after being manufactured at temperatures ranging from 250 to 350 °C. Figure 3(a), (b), and (c) depict SEM images of ZnTe thin films grown under pressure for 10 minutes at temperatures of 250, 300, and 350 °C, respectively. The SEM images reveal an uneven distribution of grains with varying sizes. The biggest grain size is visualized in the ZnTe layer grown at 300, as shown in Figure 3(b). This observation is consistent with X-ray Diffraction (XRD) results, which indicates the highest peak intensity at 300 °C in the ZnTe thin films [34]. Additionally, ZnTe thin films that are annealed at 350 °C show a smoother surface than those that are annealed at 250 and 300 °C, respectively.

## 3.3 EDS analysis of deposited ZnTe thin films

The formation analysis of ZnTe thin films is conducted through energy dispersive spectroscopy (EDS) data at various locations. Table 2 presents the atomic ratio (%) of the combined elemental composition of Zn: Te at 250, 300, and 350 °C are 87.11: 12.89, 89.68: 10.32 and 87.82: 12.18, respectively. The results reveal an excess of zinc over tellurium is produced in the ZnTe thin films prepared from a ZnTe powder using thiol-amine co-solvents. The resulting thin film is classified as n-type ZnTe due to a surplus of Zn than Te in the compound at every annealing temperature [35].

**Table 2:** The EDS elemental composition of ZnTe film annealed at various temperatures.

| Annealing Temperature | Elements | | Stoichiometric ratio [Te]/[Zn] |
|---|---|---|---|
| | Zn (at%) | Te (at%) | |
| 250 °C | 87.11 | 12.89 | 0.15 |
| 300 °C | 89.68 | 10.32 | 0.11 |
| 350 °C | 87.82 | 12.18 | 0.14 |

## 3.4 FTIR study of ZnTe thin films

The FTIR technique serves as a crucial tool for characterizing various aspects of compounds, including their chemical bonding, molecular group vibrations, rotations, and overall structural properties. Figure 4 illustrates the FTIR spectrum, showcasing three distinct ZnTe thin films annealed at varying temperatures. This spectrum covers the wavenumber range of 4000-225 cm$^{-1}$, obtained using a spectroscopic instrument.

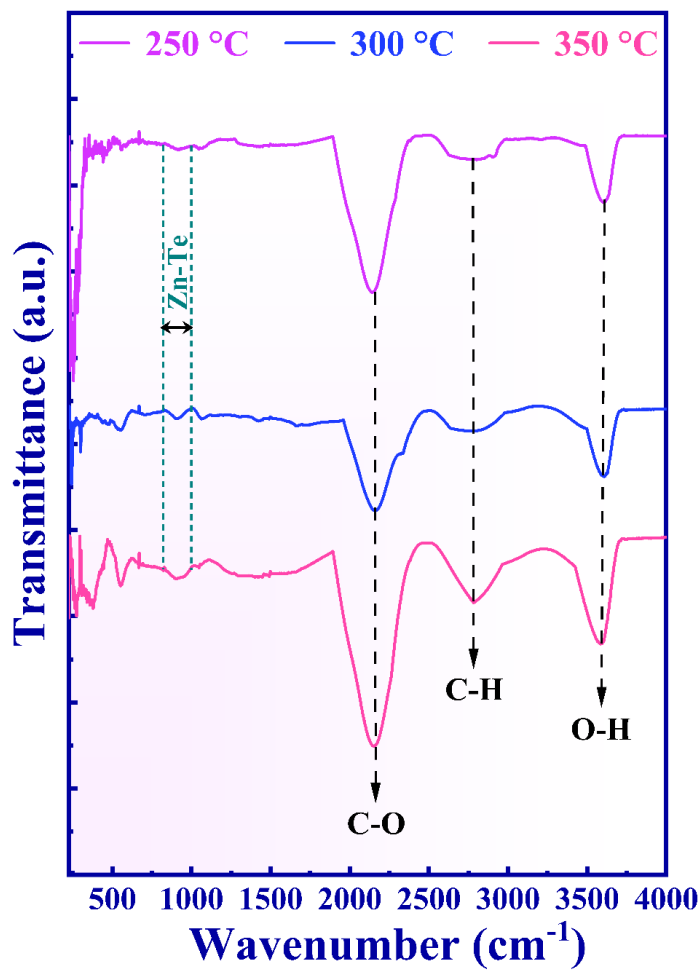

**Figure 4:** The FTIR spectrum of spin-coated ZnTe films annealed at various temperatures.

The successful synthesis of ZnTe is confirmed by the characteristic peaks observed in the infrared spectrum, typically ranging between 750-1000 cm$^{-1}$. These peaks confirm the formation of ZnTe, marking a milestone in the synthesis process [36-38]. Additionally, the peak at 2150 cm$^{-1}$ indicates the presence of CO species [39], while the peak at ~2804 cm$^{-1}$ specifies the C-H stretching vibrations bond [40]. Furthermore, broad absorption peaks noted at ~3597 cm$^{-1}$ can be linked to the O-H stretching vibrations of absorbed water on the film surface [41]. The FTIR analysis provides valuable insights into the structural and compositional characteristics of the ZnTe films, shedding light on their temperature-dependent properties and identifying key absorption features indicative of specific molecular interactions and surface phenomena.

*3.5 Optical study*

Figure 5 illustrates the fluctuations in transmittance of ZnTe thin film across an optical wavelength range of 320-1120 nm for three distinct annealing temperatures. It is evident from the observation that the transmittance of the ZnTe thin films fabricated by the spin coating method demonstrates a notable enhancement as the annealing temperature rises. Remarkably, the highest transmittance results at ~1120 nm, which is almost 70% for an annealing temperature of 350 °C. This effect can be attributed to the enhanced surface smoothness and improved crystallinity attained at higher annealing temperatures [26], which consistence with the XRD and SEM studies. It is interesting to note that transmittance consistently decreases from the infrared to the visible region, regardless of the annealing temperature. This tendency most likely results from the increased scattering losses seen in visible wavelength ranges [42].

The measurement of transmittance is important as it relates to the absorption coefficient by the following equation:

$$\alpha = \frac{ln\left(\frac{1}{T}\right)}{t} \qquad (4)$$

In the equation, T stands for the transmittance and t for the film thickness.

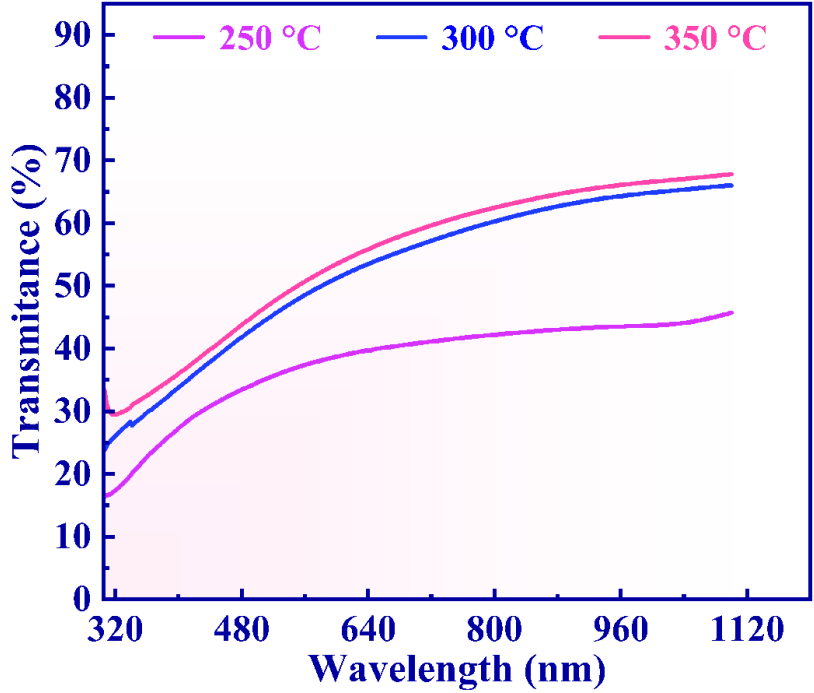

**Figure 5:** Transmittance spectra of ZnTe thin films annealed at various temperatures.

The absorption coefficient is further connected to the photon energy and this kinship can be expressed by the upcoming equation:

$$(\alpha h\nu)^2 = A(h\nu - E_g) \tag{5}$$

The above-mentioned equation is known as the Tauc equation, where A denotes a constant and $E_g$ signifies the bandgap. The linear segment of the $(\alpha h\nu)^2$ versus photon energy ($h\nu$) graph allows us to estimate the energy bandgap ($E_g$) when $(\alpha h\nu)^2$ equals zero. The Tauc plots have been visually represented by the following Figure 6 at three different annealing temperatures. The measured bandgaps from these plots are 2.92, 2.84, and 2.5 eV at 250 °C, 300 °C, and 350 °C temperatures, respectively. The figures indicate that the bandgap of ZnTe abets and reaches proximity to the bulk value as annealing temperature rises. The decline in $E_g$ can be explained by the rise in crystal size and the reduction in internal strain and dislocation density. These observations are consistent with the XRD findings and corroborate previous studies [43,44].

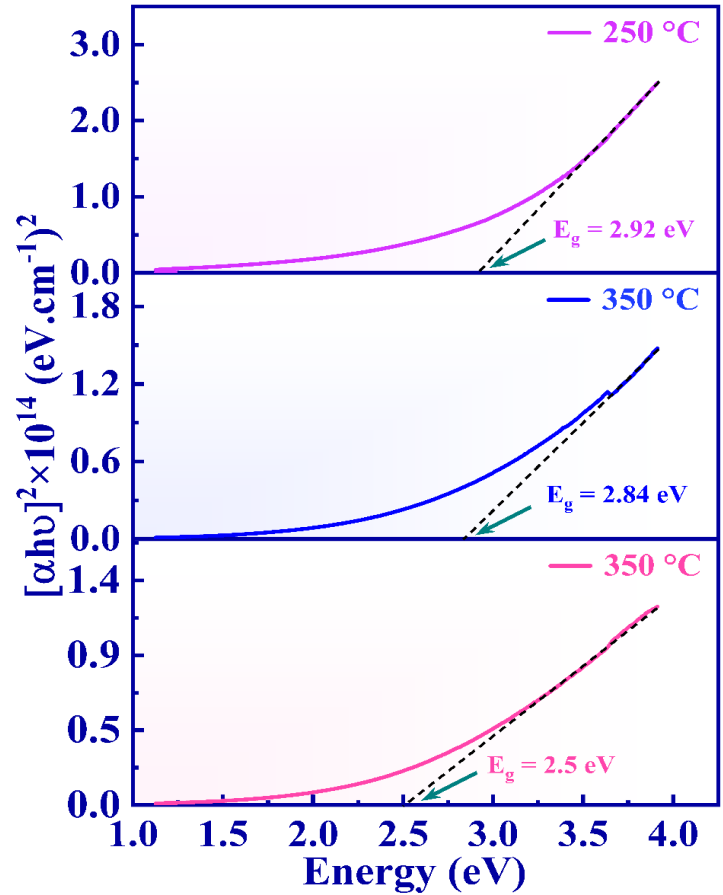

**Figure 6:** The optical bandgap of spin-coated ZnTe films annealed at various temperatures.

## 4. Conclusion

The fabrication and characterization of zinc telluride (ZnTe) thin films using a spin coating technique with ZnTe powder directly dissolving in thiol-amine co-solvents have provided valuable insights into their properties. This technique provides accurate control over film thickness, simplicity, and cost-effectiveness in contrast to conventional methods. Annealing at various conditions with stress, results in significant variations in thin film characteristics. XRD measurement confirmed the polycrystalline phase of the fabricated film. SEM analysis revealed varying grain sizes with the largest observed at 300 °C and the better smoother surface at 350 °C. Moreover, the ZnTe thin film composition is determined by analyzing EDS which confirms the n-type ZnTe thin films. FTIR spectroscopy reveals temperature-dependent molecular interactions. The ZnTe has been successfully synthesized, as indicated by the characteristic peaks that are visible in the 750-1000 cm$^{-1}$ infrared spectrum. These observable peak provide clear markers of ZnTe production, demonstrating the efficacy of the synthesis technique used. The optical study demonstrates decreasing bandgap values of 2.92, 2.84, and 2.5 eV for annealing temperatures of 250, 300, and 350 °C respectively, indicating improved structural properties. These outcomes not only justifies the synthesis procedure but also opens up new avenues for ZnTe's investigation and use in a variety of optoelectronic device applications.

**Acknowledgments**

The authors extend sincere appreciation to the Bangladesh Council of Scientific and Industrial Research (BCSIR) in Dhaka, Bangladesh, for their invaluable support and provision of facilities during XRD studies. They gratefully acknowledge the Electron Microscopy Facility at Clemson University, USA, for providing SEM and cross-sectional SEM studies. We would especially like to thank Bangladesh Council of Scientific and Industrial Research (BCSIR), Rajshahi, Bangladesh for providing EDS analysis services. Furthermore, thanks are given to the University of Rajshahi's Central Science Laboratory for their help with FTIR analysis. Additionally, the authors thank Plasma Science and Technology Lab (PSTL) for their assistance with the optical study.

***Corresponding author**

**Email:** jak_apee@ru.ac.bd (Jaker Hossain)

**Data Availability**

The data used to support the findings of this study are available from the corresponding author upon request.

**Conflicts of Interest**

The authors have no conflicts of interest.

**Declaration of generative AI and AI-assisted technologies:** The authors did not embark on AI or AI-assisted technologies in any step in preparing this manuscript.